\documentclass[pdflatex,sn-mathphys-num]{sn-jnl}
\usepackage{graphicx}%
\usepackage{multirow}%
\usepackage{amsmath,amssymb,amsfonts}%
\usepackage{amsthm}%
\usepackage{mathrsfs}%
\usepackage[title]{appendix}%
\usepackage{xcolor}%
\usepackage{textcomp}%
\usepackage{manyfoot}%
\usepackage{booktabs}%
\usepackage{algorithm}%
\usepackage{algorithmicx}%
\usepackage{algpseudocode}%
\usepackage{listings}%
\usepackage{anyfontsize}

\begin{document}

\title[Ultra-large mutually synchronized networks of 10 nm spin Hall nano-oscillators]{Ultra-large mutually synchronized networks of 10 nm spin Hall nano-oscillators}

\author*[1]{\fnm{Nilamani} \sur{Behera}}\email{nilamani.behera@physics.gu.se}
\equalcont{These authors contributed equally to this work.}
\author[1]{\fnm{Avinash~Kumar} \sur{Chaurasiya}}
\equalcont{These authors contributed equally to this work.}
\author[1,2,3]{\fnm{Akash} \sur{Kumar}}
\equalcont{These authors contributed equally to this work.}
\author[1]{\fnm{ Roman} \sur{Khymyn}}
\author[1]{\fnm{ Artem} \sur{Litvinenko}}
\author[1,4]{\fnm{Lakhan} \sur{Bainsla}}
\author[1,2,3]{\fnm{ Ahmad~A.} \sur{Awad}}
\author*[1,2,3]{\fnm{Johan} \sur{Åkerman}}\email{johan.akerman@physics.gu.se}

\affil[1]{\orgdiv{Department of Physics}, \orgname{University of Gothenburg}, \orgaddress{\street{Fysikgränd 3}, \city{ Gothenburg}, \postcode{412 96}, \country{Sweden}}}

\affil[2]{\orgdiv{Research Institute of Electrical Communication (RIEC)}, \orgname{Tohoku University}, \orgaddress{\street{2-1-1 Katahira, Aoba-ku}, \city{Sendai}, \postcode{980-8577}, \country{Japan}}}

\affil[3]{\orgdiv{Center for Science and Innovation in Spintronics (CSIS)}, \orgname{Tohoku University}, \orgaddress{\street{2-1-1 Katahira, Aoba-ku}, \city{Sendai}, \postcode{980-8577}, \state{State}, \country{Japan}}}

\affil[4]{\orgdiv{Department of Physics}, \orgname{Indian Institute of Technology Ropar}, \orgaddress{\city{Rupnagar}, \postcode{140001}, \country{India}}}

\abstract{While mutually interacting spin Hall nano-oscillators (SHNOs) hold great promise for wireless communication, neural networks, neuromorphic computing, and Ising machines, the highest number of synchronized SHNOs remains limited to $N$ = 64. Using ultra-narrow 10 and 20-nm nano-constrictions in W-Ta/CoFeB/MgO trilayers, we demonstrate mutually synchronized SHNO networks of up to $N$ 
= 105,000. The microwave power and quality factor scale as 
$N$ with new record values of 9 nW and $1.04 \times 10^6$, respectively. An unexpectedly strong array size dependence of the frequency-current tunability is explained by magnon exchange between nano-constrictions and magnon losses at the array edges, further corroborated by micromagnetic simulations and Brillouin light scattering microscopy. Our results represent a significant step towards viable SHNO network applications in wireless communication and unconventional computing. 
}
\keywords{spin Hall nano-oscillators, mutual synchronizations, magnon exchange, Brillouin light scattering microscopy}

\maketitle

\section*{Introduction}\label{sec1}

Recent technological demand for energy-efficient, spin wave-based technology is growing due to its potential for faster computation and information transmission beyond CMOS devices~\cite{chumak2022advances,gonzalez2024spintronic,flebus20242024,finocchio2024roadmap}. Spin Hall nano-oscillators (SHNOs), an emerging class of spintronic devices capable of generating propagating spin waves over long distances~\cite{Fulara2019SciAdv,kumar2025spin}, garner particular attention thanks to their unique 
properties of dimensions down to 10 nm~\cite{Demidov2014apl,durrenfeld2017nanoscale,Haidar2019natcomm,awad2020width,behera2024ultra,kumar2024mutual,montoya2025anomalous,ren2023hybrid,hache2024nanoscale}, 
strong mutual synchronization~\cite{Awad2016natphys,Zahedinejad2020natnano,kumar2023robust,kumar2025spin}, voltage tunability\cite{fulara2020natcomm,kumar2022nanoscale,choi2022voltage}, and memristive gating~\cite{zahedinejad2022memristive,khademi2023large}, making them attractive for wireless communication~\cite{ramaswamy2016wireless,sharma2021electrically}, ultra-fast spectrum analysis~\cite{louis2018ultra,litvinenko2022ultrafast,gupta2024ultra}, neural networks~\cite{barabasi1999emergence,fortunato2018science,Torrejon2017Nature,Zahedinejad2020natnano,romera2018nature,delacour2023mixed,hoppensteadt1999oscillatory}, and oscillator array-based Ising machines~\cite{houshang2022phase,litvinenko2023spinwave,graber2024integrated,albertsson2021ultrafast}. Of particular importance for these applications is the total number of mutually synchronized SHNOs ($N$) as both the microwave output power and the quality factor scale linearly with $N$~\cite{Zahedinejad2020natnano}. In addition, to solve large combinatorial problems using sparse Ising machines, one needs a very large number of interacting oscillators~\cite{Grimaldi2023prappl,albertsson2020magnetic}. 

While operational SHNO arrays of up to 100 nano-oscillators have been demonstrated, the maximum number of mutually synchronized SHNOs remains limited to 64~\cite{Zahedinejad2020natnano}. As closely spaced nano-oscillators synchronize more strongly, due to coupling mechanisms such as magneto-dipolar coupling~\cite{slavin2009nonlinear,erokhin2014robust,lebrun2017mutual}, direct exchange~\cite{Awad2016natphys}, and spin waves~\cite{slavin2006theory,chen2009phase,Kendziorczyk2016prb,kendziorczyk2014spin}, increasing the number of mutually synchronized SHNOs will require a substantial reduction in the oscillator spacing, currently at 100 nm~\cite{Zahedinejad2020natnano}. In addition, temperature gradients in large SHNO arrays are detrimental to mutual synchronization as nominally identical nano-oscillators operating at different temperatures can differ substantially in their microwave frequencies~\cite{behera2024ultra,kumar2023robust,litvinenko2023phase}. It is, therefore, important to reduce the SHNO power dissipation and maximize heat conduction away from the array.

Here, we combine the three strategies \emph{i}) reducing the inter-oscillator distance, \emph{ii}) reducing the SHNO power dissipation, and \emph{iii}) increasing the heat conduction, in an attempt to maximize the number of mutually synchronized SHNOs. As our SHNO substrate, we use native Si with excellent heat conduction, on top of which we add a 3 nm thick Al${_2}$O${_3}$ layer to electrically insulate the slightly conducting Si while retaining the heat conduction~\cite{behera2024ultra}. We use a spin-orbit-efficiency-optimized W-Ta alloy as our spin current source~\cite{behera2022energy}, a 1.4 nm thick low-damping CoFeB alloy as our ferromagnet, and a MgO capping layer to provide additional perpendicular magnetic anisotropy, 
which further reduces the threshold current density~\cite{Fulara2019SciAdv}. Using this Si/Al${_2}$O${_3}$/W-Ta/CoFeB/MgO material stack, we fabricate two large sets of different size square and rectangular SHNO arrays, from 100 to 105,000 oscillators, one based on 10-nm nano-constrictions spaced 24 nm apart, the other on 20-nm nano-constrictions with 40 nm spacing. Using electrical power spectral density measurements and scanning micro-Brillouin light scattering microscopy, we demonstrate robust mutual synchronization in all investigated arrays, evidenced by single ultrasharp microwave signals with output power and quality factors both scaling as $N$. A new record low linewidth of 25.3 kHz at an operating frequency of 26.2 GHz is observed, 
resulting in a record quality factor $Q= f / \Delta f = 1.04 \times 10^{6}$. A new record microwave output power for anisotropic magnetoresistance-based spintronic oscillators of 9 nW is observed, 
surpassing some of the literature values for magnetic tunnel junction-based oscillators.

While both the threshold current density and the microwave frequency at this threshold are independent of the number of oscillators, the current \emph{dependence} of the microwave frequency is found to be strongly dependent on array size, which is explained by a new model based on the constructive exchange of coherent propagating magnons deep within the array and magnon loss at the two array edges connected to unpatterned magnetic areas. Our very large SHNO arrays hence not only deliver several new records in terms of microwave signal quality, critically important to applications in wireless communication, ultra-fast spectrum analysis, neuromorphic computing, and Ising machines, but also open up for unexpected new physics and the possibility for network science on the nano scale~\cite{barabasi1999emergence} when the number of interacting nano-oscillators becomes very large.

\section*{Results}\label{sec2}
\subsection*{The spin Hall nano-oscillator arrays}\label{subsec2}

\begin{figure}
\includegraphics[width=\linewidth]{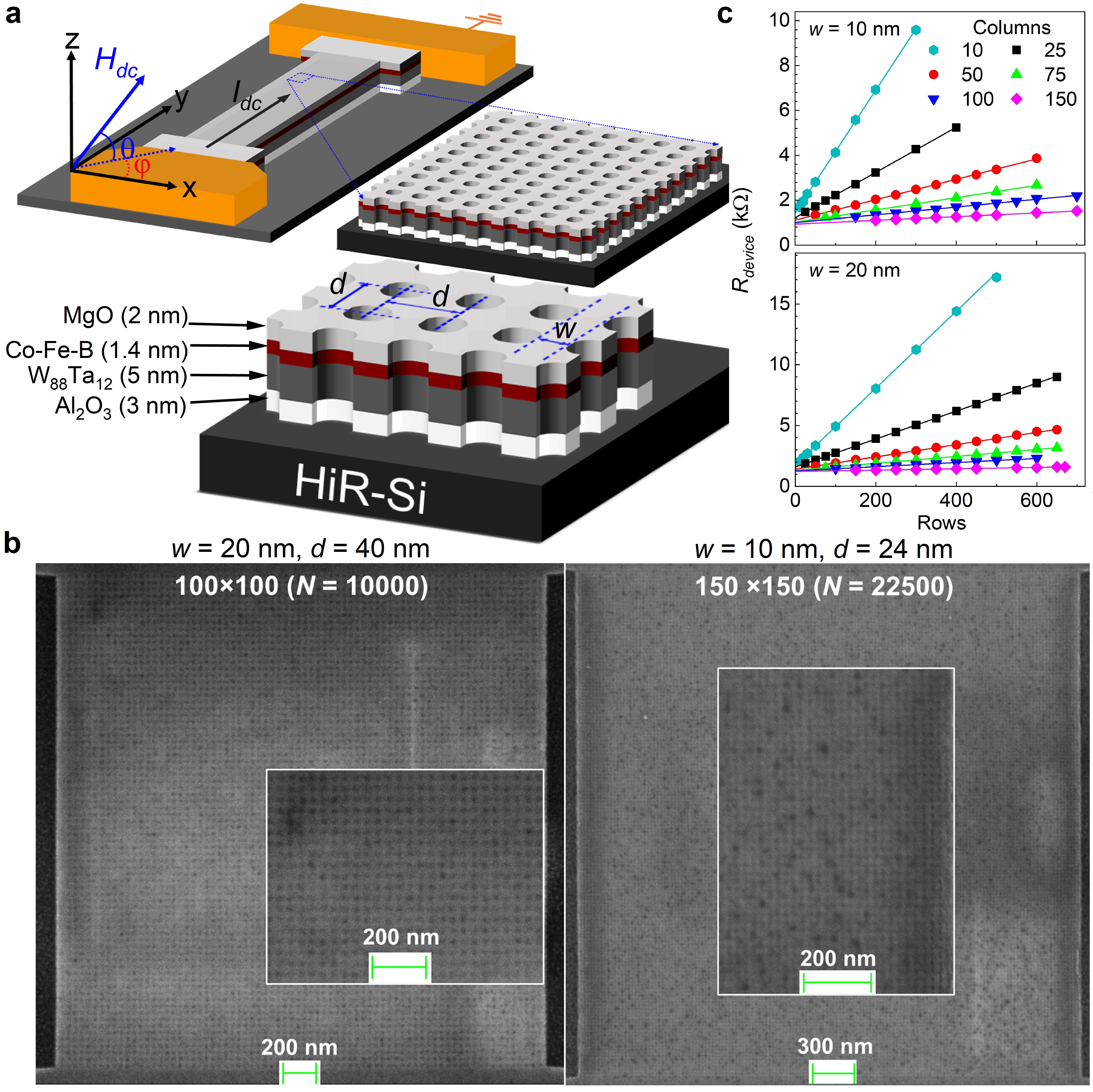}
\caption{\textbf{The SHNO arrays.}
\textbf{a,} Schematic of the SHNO arrays and their material stack, showing consecutive zoom-ins. The top cartoon shows a small part of the thick Cu/Pt contact pads (orange), the remaining part of the mesa without any nano-constrictions (light grey), and the actual nano-constriction array (darker grey). The directions of the drive current and the applied field are indicated. The bottom cartoon shows the material stack and the nano-constriction width ($w$) and center-to-center separation ($d$). \textbf{b,}  SEM images of a 100$\times$100 array made from 20-nm nano-constrictions, and a 150$\times$150 array made from 10-nm nano-constrictions.
\textbf{c,} Device resistance \emph{vs.}~number of rows for different number of columns. The solid lines are fits to Eq.1 (10-nm) and Eq.2 (20-nm) in the supplementary materials.}
\label{fig:Resistance} 
\end{figure}

Figure~\ref{fig:Resistance}a shows a schematic of a typical SHNO array device with the thick Cu/Pt contact leads (orange) connecting to a patterned mesa of the material stack (gray). The lighter gray area corresponds to regions of the mesa without nano-constrictions, while the slightly darker area indicates the actual nano-constriction array, as seen in the first zoom-in. The final zoom-in shows the Al${_2}$O${_3}$/W-Ta/CoFeB/MgO material stack and the width ($w$) and center-to-center separation ($d$) of the nano-constrictions. Figure~\ref{fig:Resistance}b shows SEM images of two square arrays: one with 100 $\times$ 100 20-nm nano-constrictions separated by 40 nm, and the other with 150 $\times$ 150 10-nm nano-constrictions separated by 24 nm. The inset SEM images, taken at higher resolution, highlight 
the good reproducibility 
and how the final dimensions closely match the nominal design values of \textit{w} and \textit{d}. As expected, the measured device resistance (Fig.~\ref{fig:Resistance}c)  scales linearly with the number of rows ($X$) and inversely with the number of columns ($Y$). All resistance values are well fitted by a model accounting for the varying dimensions of the remaining mesa (see supplementary, Eqs.~1\&2), with the individual resistances of the 10-nm and 20-nm nanoconstrictions being 293 $\Omega$ and 348 $\Omega$ independent of array size (Fig.S2 in the Supplementary Information). The anisotropic magnetoresistance (AMR) is $\sim$0.9 $\%$ for both widths and all arrays (Fig.S3 in the Supplementary Information). 

\subsection*{Auto-oscillations \emph{vs.}~array size}\label{subsec3}

\begin{figure}
\centering
\includegraphics[width=11cm]{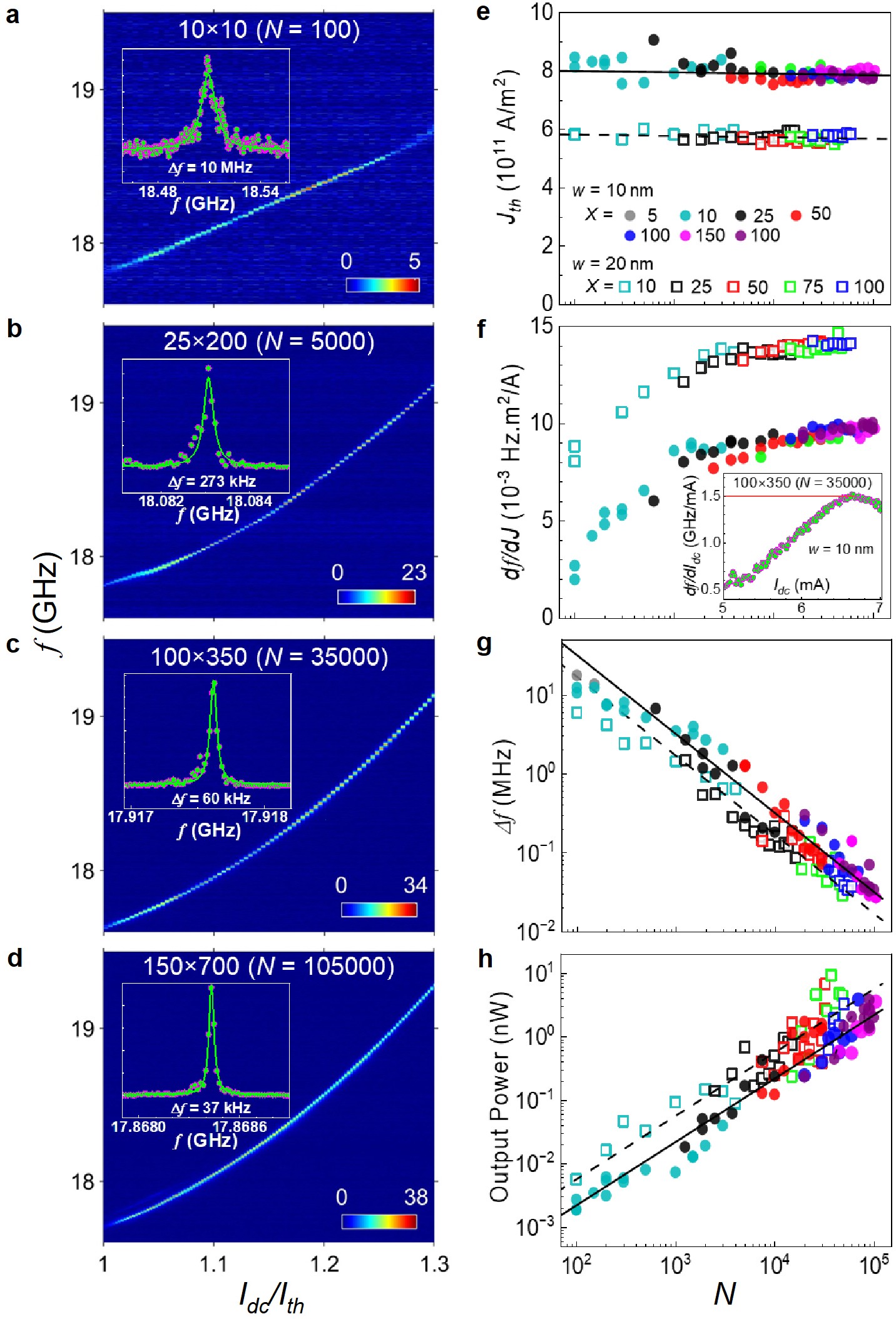}
\caption{\textbf{Auto-oscillation power spectral density (PSD), threshold current, tunability, linewidth, and power.}
\textbf{a,b,c,d,} PSD \emph{vs.}~criticality ($I_{dc}/I_{th}$) for four representative arrays with \textit{N}~=~100,~5000,~35000, and 105000 nano-constrictions ($w=$ 10-nm). The color bars indicate the peak power in dB over the noise floor. The insets show individual spectra around the lowest measured linewidth ($\Delta f$) for each array, together with Lorentzian fits. \textbf{e,} Threshold current, \textbf{f,} maximum frequency tunability, \textbf{g,} linewidth, and \textbf{h,} total microwave power, \emph{vs.}~number of SHNOs ($N$). The inset in \textbf{(f)} shows how the maximum frequency tunability is extracted from a plot of $df/dI_{dc}$. Solid ($w=$ 10-nm) and dashed ($w=$ 20-nm) lines are fits to $N^{-1}$ in \textbf{(g)} and to $N$ in \textbf{(h)}.  
}
\label{fig:PSDs} 

\end{figure}

Auto-oscillations on 
well-defined single microwave signals---consistent with complete mutual synchronization---were observed in all investigated arrays, with only a few arrays showing multiple signals just above auto-oscillation onset due to partial synchronization. Fig.~\ref{fig:PSDs}a--d show the microwave power spectral density (PSD) \emph{vs.}~normalized current ($I_{dc}/I_{th}$) of four representative 10-nm nano-constriction arrays spanning the full array size range from 10~$\times$~10 to 150~$\times$~700 nano-constrictions. Additional PSDs of 10-nm and 20-nm large SHNO arrays are shown in the Supplementary Information file,~2.1~\&~ 2.2~(Figs.~S4~\&~S5).  
The auto-oscillation threshold current ($I_{th}$) is independent of the number of rows and linear in the number of columns (Fig.~S6, section~2.3 in the Supplementary Information) such that the threshold current density ($J_{th}$) is independent of array size and only slightly different between the 10-nm and 20-nm arrays (Fig.~\ref{fig:PSDs}e), corroborating the excellent reproducibility of the nanolithography process throughout all arrays; the auto-oscillation frequency at the threshold is similarly independent of array size. 

However, as can be seen in Fig.~\ref{fig:PSDs}a--d, the positive and monotonic current \emph{dependence} of the auto-oscillation frequency increases substantially with array size, showing an essentially linear and weak current dependence in the smallest arrays while becoming increasingly strong and quadratic in the larger arrays. In the inset of Fig.~\ref{fig:PSDs}f, we plot the derivative of the frequency \emph{w.r.t}~current for a large array, showing a linear behavior 
up to a maximum, $\mathit{(df/dI_{dc})_{max}}$, which we then plot \emph{vs.}~$N$ for all arrays in the main part of Fig.~\ref{fig:PSDs}f. The strong dependence of the current tunability---about a factor of five for the 10-nm arrays---scales surprisingly well with the total number of nano-constrictions with, possibly, a very weak additional dependence on the number of columns. 

For wireless communication applications, mutual synchronization offers the two equally important benefits of reducing the auto-oscillation linewidth ($\Delta f$) and increasing the total microwave power. 
In Fig.~\ref{fig:PSDs}g, we plot the lowest measured $\Delta f$ \emph{vs.}~$N$ for all the investigated arrays, with $\Delta f$ extracted from Lorentzian fits to the type of spectra shown in the insets of Fig.~\ref{fig:PSDs}a--d. The solid ($w=$ 10-nm) and dashed ($w=$ 20-nm) lines are fits to $N^{-1}$, which is the theoretically expected dependence. As expected from theory, the linewidth of the 20-nm nano-constriction arrays is considerably better than that of the 10-nm nano-constriction arrays
, since the mode volume and energy scales approximately as $w^2$. From the excellent scaling, we conclude that mutual synchronization is achieved in all arrays, with 
one of the largest 20-nm arrays ($N=75 \times 650=48,750$ nano-oscillators) showing the best linewidth of all arrays. Additional microwave signal analysis on this array over a wider range of magnetic field and current resulted in a record low linewidth of 25.3 kHz at an auto-oscillation frequency of 26.2 GHz, corresponding to a new record-high quality factor for any spintronic oscillator of $Q=f/\Delta f=1.04 \times 10^6$ (Fig.~S8). 

In Fig.~\ref{fig:PSDs}h, we plot the highest generated microwave output power \emph{vs.}~$N$ of each array together with linear fits, again showing excellent scaling. For the 10-nm arrays, the output power reaches a maximum of $\sim$4 nW in the 105,000 SHNO array. The highest output microwave power observed in any array is $\sim$9 nW for a 20-nm 75 $\times$ 500 array with 37,500 SHNOs. 
These observed output microwave power values are the highest to date among all reports on SHNOs and even surpass some of the literature values achieved with magnetic tunnel junction based spin torque nano-oscillators.

\subsection*{Brillouin light scattering microscopy}\label{subsec4}

\begin{figure}
\centering
\includegraphics[width=13cm]{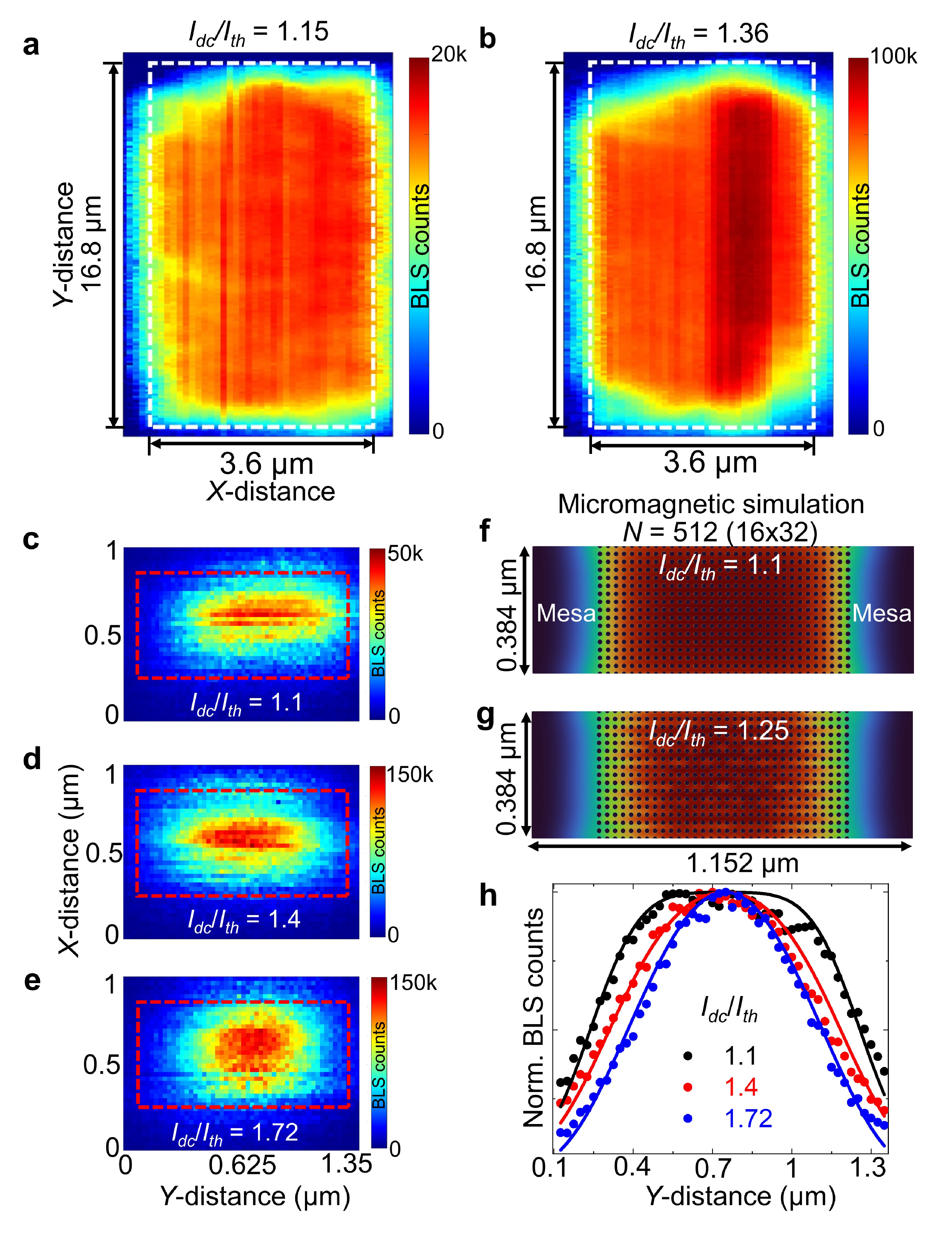}
\caption{\textbf{Micro-BLS microscopy and micromagnetic simulation.}
\textbf{a,b,} Micro-BLS microscopy images (log scale) of the largest array with 105,000 10-nm SHNOs at \textbf{(a)} $I_{dc}=$~6.5 mA, and \textbf{(b)} $I_{dc}=$~7.6 mA. The white dashed box indicates the extent of the array. \textbf{c,d,e,} BLS counts (linear scale) within a 25$\times$50 10-nm array at increasing currents: $I_{dc}=$~1.03, 1.3, and 1.6 mA. The red dashed box indicates the extent of the array.  \textbf{f,g,} Micromagnetic simulations of a 16$\times$32 10-nm SHNO array at two different currents. \textbf{h,} Normalized BLS counts in \textbf{(c-e)} \emph{vs.}~Y position together with fits to a model described in the text.}
\label{fig:BLSMaps} 
\end{figure}

To directly visualize the auto-oscillations and the mutual synchronization, we use scanning micro-Brillouin light scattering ($\mu$-BLS) microscopy and map out the spin wave intensity inside and outside of the arrays. 
Fig.~\ref{fig:BLSMaps}a--b show BLS maps of the largest 105,000-SHNO array taken at two current levels, one just above threshold and the other 36\% above threshold, where the higher current shows a five-fold increase in the maximum spin wave intensity. Close to threshold, most of the array shows a relatively uniform spin wave intensity with some fall-off towards the shorter edges. 
A significant fraction of this fall-off can be ascribed to the diffraction limited Gaussian beam profile of the BLS laser, which adds $\sim$300 nm of Gaussian blur and limits the spatial resolution, most easily seen along the four array edges. However, at the higher current (Fig.~\ref{fig:BLSMaps}b), the fall-off becomes even more pronounced at the short edges, predominantly so towards their corners, whereas there are no corresponding changes at the long edges. 

To study this unexpected behavior in more detail, and allow for a direct comparison with micromagnetic simulations, we acquired BLS maps at increasing current for a 
25 $\times$ 50 10-nm array, where the edge regions represent a larger fraction of the auto-oscillating area (Fig.~\ref{fig:BLSMaps}c--e); the three current values were chosen as 10\%, 40\%, and 72\% above threshold. As can be seen, the region of highest spin wave intensity gradually shrinks towards the center of the array, away from the short edges. We quantify this behavior in Fig.~\ref{fig:BLSMaps}h, where the data points correspond to the normalized BLS counts \emph{vs.}~position along the Y axis of the array and the solid lines are fits to a model discussed below. For increasing current, the relative spin wave intensity in the array clearly gravitates towards the center. 

\subsection*{Micromagnetic simulations and model}\label{subsec5}

We then carried out micromagnetic simulations on similarly sized 16 $\times$ 32 arrays with the same 10-nm nano-constriction width, 24-nm separation, and using the same material and magnetic field parameters as in the experiment (Fig.~\ref{fig:BLSMaps}f--g). We reproduce the propagating nature of the auto-oscillating spin wave modes, and, more importantly, also reproduce the non-uniform spin wave intensity towards the short edges, including the current-dependent focusing of the spin wave intensity towards the array center. It is noteworthy that the simulations, unaffected by the Gaussian blur of the BLS microscope, only show a non-uniform spin wave intensity towards the short array edges. At the higher current, streaks of variable spin wave intensity on a length scale much below the spatial resolution of the BLS microscope also appear.

The non-uniform spin wave intensity can be explained by a simple model (see Supplementary Information, Section 3.2) where each nano-constriction emits propagating spin waves into its surroundings. Deep within the array, far away from the short edges, all nano-constrictions exchange spin waves in equal proportions with its surrounding neighbors, resulting in a uniform spin wave intensity. However, at the short edges, spin waves are emitted into the unpatterned, non-auto-oscillating mesa, where they rapidly dissipate. The edge nano-constrictions hence exhibit a much lower amplitude of their magnetization precession due to the loss of magnons into the mesa and, therefore, also emit much fewer spin waves into their nearest nano-constriction neighbors. Consequently, the magnon loss into the mesa is felt well into the array over a characteristic length given by the magnon decay length.

The model can also explain the unexpected gradual increase in the current tunability of the auto-oscillation frequency as $N$ increases (Fig.~\ref{fig:PSDs}f). At the auto-oscillation threshold ($\mathit{J_{th}}$), each nano-constriction experiences approximately the same spin wave losses, regardless of its position in the array, consistent with the independence of $\mathit{J_{th}}$ on $N$. However, above $\mathit{J_{th}}$, and assuming coherent mutual synchronization, the net radiative losses experienced by a nano-constriction deep inside the array will decrease with increasing auto-oscillation amplitude thanks to the coherent spin waves it receives from its neighbors. To first approximation, this reduction is linear in $\mathit{J_{dc}}$, which transforms the linear $f\sim \mathit{J_{dc}}$ dependence in single SHNOs or small arrays into an increasingly quadratic $f\sim \mathit{J_{dc}}^2$ dependence in the larger arrays. The larger the array, the larger the effect, up to a characteristic length scale, again given by a spin wave decay length.  

\section*{Conclusion}\label{sec3}

Our demonstration of networks with up to 105,000 mutually synchronized nano-oscillators extends the previous record by over three orders of magnitude and ushers in a new era where highly scalable and energy-efficient nano-oscillator networks can be used in a wide range of applications. Our largest networks demonstrate record values for microwave output power and signal quality, directly relevant for wireless communication, signal processing, and ultrafast spectral analysis; in Supplementary Information, Fig.~S9, we benchmark these results to all values reported in the literature.  
The ultra-large networks will significantly benefit neuromorphic computing, reservoir computing, and Ising machines, and may also be used for fundamental and applied studies in network science. Through their unique combination of high frequency tunability, substantial non-linearity
, and phase-tunable mutual synchronization, these large SHNO networks offer a versatile and efficient platform for a wide range of 
complex tasks such as speech recognition, image classification, and predictive modeling for chaotic time-series data, along with tasks requiring transformation and forecasting in time-series predictions~\cite{Torrejon2017Nature, romera2018nature}. The field and current tunable non-linearity supports adaptive transformation in time-series data, allowing for task-specific configurations, ideal for task-adaptive physical reservoirs~\cite{lee2024task,ng2024retinomorphic}. It is noteworthy that all our arrays exhibited robust mutual synchronization, suggesting that we have not yet reached an upper network size limit. Scaling beyond millions of SHNOs could hence be possible by increasing the coupling strength through further reduction of the separation and increasing the magnetization and thickness of the ferromagnetic layer.

\section*{Methods}\label{sec4}
\subsection*{Material growth}\label{subsec6}
Al${_2}$O${_3}$(3 nm)/W$_{88}$Ta$_{12}$(5 nm)/Co$_{20}$Fe$_{60}$B$_{20}$(1.4 nm)/MgO(2 nm) material stacks were magnetron sputtered onto high-resistance intrinsic (undoped) Si substrates. All sputtering parameters were the same as in previous works where the W-Ta alloy and the Al${_2}$O${_3}$ seed layer were optimized for minimal heat generation, maximum heat conduction and lower current shunting~\cite{behera2022energy, behera2024ultra}.

\subsection*{Design and fabrication of square and rectangular SHNO arrays}\label{subsec11}
Nano-constriction spin Hall nano-oscillators (SHNOs) of two widths, $w=$ 10 and 20-nm, were fabricated using E-beam lithography (EBL). 
146 different square and rectangular SHNO arrays with different number of columns (X = 10--150) and rows (Y = 10--1000) were defined in the center of 6 $\times$ 22$~\mu$m$^2$ mesas for the 10-nm SHNOs and 8 $\times$ 30$~\mu$m$^2$ mesas for the 20-nm SHNOs. 
The center-to-center separation of the 10-nm SHNOs  
was $d=$ 24 nm, while that of the 20-nm SHNOS 
was $d=$ 40 nm.  
Between the SHNO arrays, we also fabricated micro-bars of three different sizes (6$\times$12, 6$\times$18, 6$\times$22$~\mu$m$^2$) for SOT characterization using ST-FMR.

EBL proceeded by covering the material stack with  
negative resist 
(HSQ, XR-1541-002, 2\%) followed by EBL exposure (Raith EBPG 5200). Ar-ion beam etching was done using an Oxford Ionfab 300 Plus etcher. 
An optical lithography lift-off process defined ground-signal-ground (GSG) co-planar waveguides (CPWs) in a sputter-deposited Cu (800 nm)/Pt (20 nm) bilayer. A Zeiss Supra 60 VP-EDX scanning electron microscope (SEM) was used to inspect the quality and dimensions of all SHNOs and arrays.

Figure~S1 shows SEM images of \textbf{(a)} 20-nm SHNO arrays with X $\times$ Y being 25$\times$25, 50$\times$50, 75$\times$75 and \textbf{(b)} 10-nm SHNO arrays with X$\times$Y being 25$\times$150, 50$\times$75, 100$\times$100. 
The insets show parts of the same devices at higher magnification. 
 
\subsection*{Electrical characterization}\label{subsec7}

The resistance and anisotropic magneto-resistance (AMR) of all SHNO arrays 
were extracted from  
resistance measurements in a vector magnet with a rotatable in-plane field of 0.25 T. 
An in-plane field angle $\phi$ = 0$^{\circ}$ was defined as the current and magnetic field directions being perpendicular (minimum resistance). 

Magnetization auto-oscillation measurements were carried out using a custom-built probe station. The device under test is mounted inside an electromagnet on a holder that allows for variable in-plane and out-of-plane field angles. Ground-signal-ground (GSG) probes connect to the corresponding GSG pads of the device and a direct current is fed to the device through the DC port of a high-frequency bias-T. A 31 dB low-noise microwave amplifier connected to the high-frequency output of the bias-T amplifies the auto-oscillation microwave voltage before it is recorded by a Rohde $\&$ Schwarz FSV40 spectrum analyzer. 

\subsection*{Microfocused-Brillouin light scattering ($\mu$-BLS) measurement}\label{subsec8}
 
A monochromatic continuous wave (CW) laser (wavelength, ($\lambda$ = 532 nm; laser power, $P_{power}$ = 0.2 mW) was focused onto the samples by a $\times$100 microscope objective (MO) with a large numerical aperture (NA = 0.75) down to a 300 nm diffraction limited spot diameter. The external magnetic field was applied at an in-plane angle ($\varphi$) of 22$^{\circ}$ 
and OOP angle ($\theta$) of 60$^{\circ}$, and a constant magnetic field magnitude of 0.5 T. The inelastically scattered light from the sample was collected by the same MO and analyzed with a Sandercock-type six-pass tandem Fabry-Perot interferometer TFP-1 (JRS Scientific Instruments). A stabilization software based on an active feedback algorithm (THATec Innovation) was employed to get long-term spatial stability during the $\mu$-BLS measurement. The experimentally obtained BLS intensity is proportional to the square of the amplitude of the dynamic magnetization.   

\backmatter

\bmhead{Supplementary information}
The additional supplementary data in support of main text are available in the Supplementary Information section.

\section*{Declarations}
\bmhead{Acknowledgements}
This work was partially supported by the Horizon 2020 research and innovation program No. 835068 "TOPSPIN". This work was also partially supported by the Swedish Research Council (VR Grant No. 2016-05980) and the Knut and Alice Wallenberg Foundation. This work was performed in part at Myfab Chalmers.

\bmhead{Author contribution}
J.{\AA}. $\&$ N.B. initiated the project. N.B., A.K.C., A.K., A.A. and J.Å. designed the experiments. N.B. optimized and prepared the thin film stacks and designed and fabricated the devices. N.B. carried out all electrical microwave, magneto-transport measurements and data analysis in discussion with J.{\AA}, A.K. and A.L. N.B. performed SEM measurements with assistance from L.B. J.{\AA}., N.B. \& A.K. were involved in the device resistance data analysis. A.K.C. performed all micro-BLS measurements and analysis in discussion with A.A. \& J.{\AA}. R.K. performed the micromagnetic simulations and developed the theoretical model. All authors contributed to the  data analysis. J.{\AA}. coordinated and supervised the project. All authors participated in the discussion and co-wrote the manuscript. 

\bmhead{Competing interests}
 The authors declare that they have no competing interests.

\bmhead{Data availability}
All data in the main text and the supplementary materials are available from the corresponding author upon reasonable request.

\bmhead{Code availability}
The codes for micromagnetic simulation and model are available from corresponding author upon reasonable request.

\bibliography{ref}

\end{document}